\documentclass[%
 reprint,
 showpacs,
 amsmath,amssymb,
 aps,
 pra,
]{revtex4-1}

\usepackage[ascii]{inputenc}
\usepackage[T1]{fontenc}
\usepackage[english]{babel}
\usepackage{graphicx}
\usepackage{microtype}
\usepackage[pdftex,colorlinks]{hyperref}
\hypersetup{%
  linkcolor=blue,%
  citecolor=blue,%
  urlcolor=blue,%
  pdftitle={Balanced gain and loss in Bose-Einstein condensates without PT
  symmetry},%
  pdfauthor={Philipp Lunt, Daniel Haag, Dennis Dast,
    Holger Cartarius, G\"unter Wunner}%
}

\DeclareMathOperator{\imag}{Im}

\newcommand{\PT}{\mathcal{PT}}

\newcommand{\mrm}{\mathrm}
\newcommand{\rmi}{\mathrm{i}}
\newcommand{\rme}{\mathrm{e}}

\begin{document}

\title{Balanced gain and loss in Bose-Einstein condensates without $\PT$
symmetry}

\author{Philipp Lunt}

\author{Daniel Haag}
\email[]{daniel.haag@itp1.uni-stuttgart.de}

\author{Dennis Dast}

\author{Holger Cartarius}

\author{G\"unter Wunner}

\affiliation{Institut f\"ur Theoretische Physik 1,
  Universit\"at Stuttgart, 70550 Stuttgart, Germany}

\date{\today}

\begin{abstract}
  Balanced gain and loss renders the mean-field description of Bose-Einstein
  condensates $\PT$ symmetric.
  However, any experimental realization has to deal with unbalancing in the
  gain and loss contributions breaking the $\PT$ symmetry.
  We will show that such an asymmetry does not necessarily lead to a system
  without a stable mean-field ground state.
  Indeed, by exploiting the nonlinear properties of the condensate, a small
  asymmetry can stabilize the system even further due to a self-regulation of
  the particle number.
\end{abstract}

\maketitle

\section{Introduction}

Since the discovery of real eigenvalues in a non-Hermitian $\PT$-symmetric
Hamiltonian by Bender and Boettcher in 1998~\cite{Bender98a} a lot of work was
put into replacing the usual concept of Hermitian quantum mechanics with the
more general condition of $\PT$
symmetry~\cite{Bender02a,Mostafazadeh08a,Mostafazadeh10a}.
In the course of the search for experimental realizations, the attention
shifted to optical systems, where $\PT$ symmetry is accomplished by a positive
and negative imaginary refractive index that in the equations effectively
models a gain and loss of the field strength~\cite{El-Ganainy07a, Klaiman08a,
  Musslimani08a, Makris08a, Makris10a}.

In fact the first realizations succeeded in two coupled optical waveguides.
In the first experiment two different absorption strengths were used to create
passive $\PT$ symmetry~\cite{Guo09a}, whereas in a subsequent realization one
waveguide was actively pumped to amplify the field strength~\cite{Ruter10a}.
This showed that while the original concepts of $\PT$ symmetry focused on
fundamental changes in the nature of quantum mechanics, its first realizations
succeeded in an effective mean-field description, which again attracted further
theoretical and experimental efforts~\cite{Peng14b, Regensburger12a}.
Another approach towards a realization lies in purely electronic
frameworks~\cite{Schindler11a, Bender13a}.

With the success of $\PT$ symmetry in these mean-field systems in mind
it is quite comprehensible that Bose-Einstein condensates should also qualify
for a realization~\cite{Klaiman08a}.
In this many-particle system, the in- and out-coupling acts directly on the
particle density, increasing or decreasing the number of particles.
This interpretation of particle loss and gain recently lead to the first real
quantum simulation of a $\PT$-symmetric system using a $^6$Li Fermi
gas~\cite{Li16a}.
Numerical calculations in spatially extended potentials confirm that
condensates are in principle able to provide all the effects known from linear
optical realizations~\cite{Cartarius12a, Cartarius12b, Dast13a, Dast13b,
Haag14b}.
Proposals for an experimental realization in analogy to two optical waveguides
have been made.
They include embedding a double-well system in a longer chain of wells with
time-dependent coupling parameters~\cite{Kreibich13a}, and the description of
two separate condensates exchanging their particles~\cite{Single14a}.
Furthermore, particle gain and loss can be realized by coupling particles
into and out from the surrounding environment.
Both processes have already been realized experimentally:
Out-coupling by a focused electron beam~\cite{Gericke08a} and in-coupling by
letting atoms fall into the condensate from a second
condensate~\cite{Robins08a}.

In real systems a perfect control of the in- and out-coupling of particles is
not possible.
Therefore, asymmetries in the imaginary potential have to be expected, i.e.,
the system is not exactly $\PT$ symmetric.
Even though $\PT$ symmetry is neither a necessary nor a sufficient condition
for real eigenvalues~\cite{Mostafazadeh10a} and its typical properties are also
found in other systems~\cite{Cannata98a, Bagchi00a}, it is not to be expected
that such a perturbation leaves the stationary states intact.
However, we will show that by increasing the in- and out-coupling parameter to
a specific strength, one can restore a single real eigenvalue.

This paper is organized as follows.
We start with a two-mode approximation, analyzing its eigenvalues and stability
in Sec.~\ref{sec:twoMode}.
Its dynamical properties and a comparison with the $\PT$-symmetric double-well
system are investigated in Sec.~\ref{sec:twoModeDyn}.
Afterwards, the discussion is extended to a spatially extended potential in
Sec.~\ref{sec:extended}.

\section{Stationary solutions}
\label{sec:twoMode}
A Bose-Einstein condensate allowing for an asymmetry in the gain and loss of
particles can be described by the Hamiltonian
\begin{equation}
  H = 
    \begin{pmatrix}
     \rmi \gamma(1+a_I) & -1\\
    -1 & -\rmi \gamma(1-a_I) 
    \end{pmatrix},
  \label{eq:AsymHamiltonian}
\end{equation}
where $a_I \in \mathbb{R}$ is the asymmetry between gain and loss and $\gamma$
is the dimensionless overall strength of the in- and out-coupling.
The relative particle loss in the second well reads $2\gamma/\tau$, where the
timescale $\tau$ is fixed by the size and shape of a trapping potential.
Using the double-well experiment of~\cite{Gati07a} as an example, an
approximate timescale of $\tau \approx 30ms$ is found.
Comparing this timescale to the losses realized in~\cite{Ott13a} shows, that
such particle losses are well within experimental possibilities.

For $a_I = 0$, the $\PT$-symmetric two-mode model~\cite{Graefe12a,Cartarius12a,
Dast13a} is restored, where in the first well particles are injected into the
system and particles are removed from the second well.
In this system the ground and first excited state are $\PT$ symmetric and
have real eigenvalues up to an exceptional point, at which eigenstates and
eigenvalues coalesce and vanish.
From this point two $\PT$-broken states with complex conjugate eigenvalues
emerge.

This behavior changes drastically for $a_I \neq 0$.
While the eigenvectors are not influenced by the asymmetry in the imaginary
part, the eigenvalues of the Hamiltonian~\eqref{eq:AsymHamiltonian} now read
\begin{equation}
  \mu_\pm = \rmi \gamma a_I \pm \sqrt{1-\gamma^2}.
  \label{eq:AsymEigenvalues}
\end{equation}
Even though the second part of this equation shows the described behavior from
the $\PT$-symmetric case, the asymmetry $a_I$ generates a purely imaginary
shift rendering both eigenvalues complex, i.e., they are no longer real
stationary solutions of the Schr\"odinger equation.

One can easily check that one of the eigenvalues reaches $\imag\mu = 0$ at
a specific parameter value $\gamma_0 = \sqrt{1/(1-a_I^2)}$, thus, the
corresponding eigenvector becomes stationary.
We stress that at this intersection point $\gamma_0$ the other eigenstate of
the two-mode system has the eigenvalue $\mu = 2\rmi a_I$ which
means that it grows exponentially for $a_I > 0$, for which the particle gain
is stronger, and decays for $a_I < 0$.
The actual eigenstate is therefore stable only if an asymmetry is chosen in
such a way that the particle loss is stronger than the gain.
It can be shown that this requirement is, in fact, mandatory for spatially
extended systems, since a stronger particle gain would enhance any high-energy
perturbations that are equally strong in both wells.

However, a major problem remains.
The actual stationary state exists only at one specific parameter $\gamma_0$
for a given asymmetry $a_I$.
This leaves the experimental realization with the same problem as before since
a small deviation from the desired gain or loss of particles could force the
particle number to grow exponentially.
However, the problem can be overcome by the particle-particle interactions
present in a Bose-Einstein condensate if they are manufactured in such a way
that a growth or decay in the particle number stabilizes the state.

\begin{figure}
  \centering
  \includegraphics{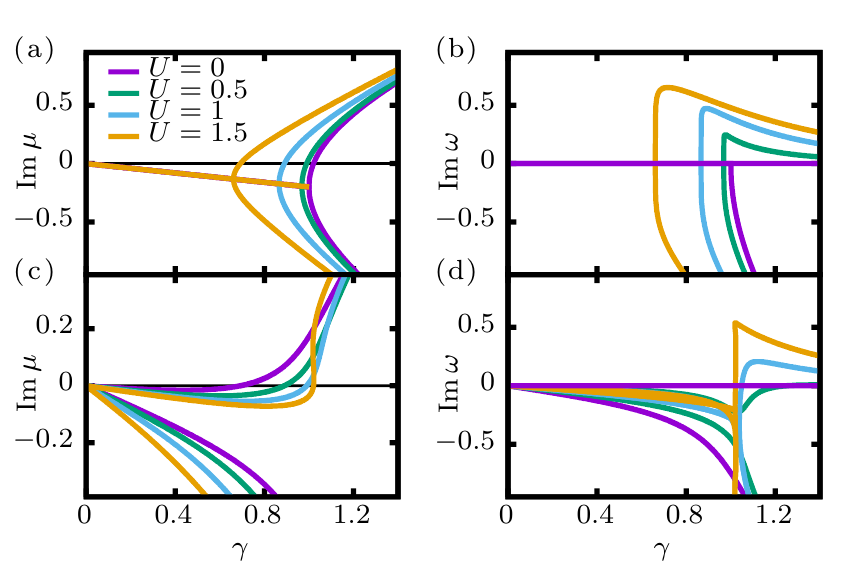}
  \caption{
    The imaginary part of the nonlinear spectrum (a) of the asymmetric double
    well~\eqref{eq:AsymHamiltonian} with $a_I = -0.2$.
    For \emph{stronger} nonlinearity parameters $U$ the upper branches
    intersects the axis $\imag\mu = 0$ at \emph{lower} parameters
    $\gamma$.
    The imaginary part of the Bogoliubov-de Gennes eigenvalues $\omega$ (b) are
    given for the upper branch, which contains the stationary state.
    In the nonlinear case this state is unstable.
    For the system described by Eq.~\eqref{eq:TwoAsymHamiltonian} with $a_R =
    -0.15$ (c), i.e., the case of asymmetric on-site energies, this behavior
    changes.
    \emph{Stronger} nonlinearity parameters $U$ now lead to upper branches that
    intersect the axis $\imag\mu = 0$ at \emph{larger} parameters
    $\gamma$.
    This holds up to $U \approx 1.5$.
    For even stronger nonlinearities, the intersection with the axis again
    moves to lower values of $\gamma$.
    The appropriate stability eigenvalues of the Bogoliubov-de Gennes equations
    (d) are now negative imaginary numbers, which shows that the stationary
    state is a dynamical attractor.
  }
  \label{fig:twoModeSpec}    
\end{figure}

Introducing a nonlinear contact interaction term leads to the two-mode
Gross-Pitaevskii equation,
\begin{equation}
  \sum\limits_{j=1}^{2} H_{ij} \psi_j + U \left| \psi_i \right|^2
  \psi_i  = \mu \psi_i,
  \label{eq:TwoModeGpe}
\end{equation}%
where $U$ specifies the strength of the interaction, which can take values from
$0$ to $2000$~\cite{Bloch05a}.
Such a contribution renders the spectrum dependent on the norm of the wave
function.
Thus, the parameter $\gamma$, for which the stationary state is found, changes
with a growth or decay of the wave function.
In the $\PT$-symmetric case, $a_I = 0$, analytical solutions were given by
Graefe et al.~\cite{Graefe12a}.
With $U > 0$, the $\PT$-broken states no longer bifurcate at $\gamma = 1$,
at which the $\PT$-symmetric states vanish.
Instead, they emerge at an earlier point, $\gamma = \sqrt{1-U^2/4}$ from the
excited state.
It is a solid hypothesis that a similar effect can be expected for $a_I < 0$,
where alongside the $\PT$-broken states, also the intersection point $\gamma_0$
moves to lower values.
This is confirmed by Fig.~\ref{fig:twoModeSpec}(a), which shows the eigenvalue
spectrum for the nonlinearity parameters $U = 0, 0.5, 1$ and $1.5$.
In these calculations, the asymmetry $a_I = -0.2$ was used.

Not only is the aforementioned hypothesis confirmed but the results already
show that the system cannot be stable:
Consider an experimental system in which the non-Hermiticity or gain-loss
parameter $\gamma$ is chosen larger than $\gamma_0$.
The ground state with $\imag\mu > 0$ is then growing instead of staying
stationary.
This effectively increases the nonlinearity and shifts the intersection point
of $\gamma$, at which the true stationary state with $\imag \mu = 0$ can
be found, to even lower values.
The distance between the chosen and the correct value of $\gamma$ increases,
and thus the error is amplifying itself due to the nonlinearity.

This instability can also be shown numerically using the corresponding
Bogoliubov-de Gennes equations~\cite{deGennes89a}.
The eigenvalues of this system of equations describe how small perturbations
of the eigenstate behave.
Positive imaginary parts show an unstable behavior, while negative imaginary
parts characterize a dynamical attractor.
These eigenvalues are shown in Fig.~\ref{fig:twoModeSpec}(b) for the
$\PT$-broken state that intersects with $\imag \mu = 0$ at $\gamma_0$.
In agreement with our prediction there is a positive imaginary part and the
states are unstable.

The influence of the nonlinearity in this simple model not only fails to
stabilize the stationary state against small deviations from $\gamma_0$ but
introduces an instability against norm oscillations.
To get rid of this additional instability the opposite behavior is required:
A stronger nonlinearity must reduce the imaginary part of the chemical
potential.
Therefore, we have to invert the movement of the intersection point $\gamma_0$
in the spectrum such that it is shifted to higher parameters $\gamma$ if the
interaction is increased, and vice versa.

To influence the overall form of the spectrum a new parameter has to be
introduced.
The inversion of the movement of the intersection point can be achieved by
introducing an additional asymmetry, now in the real part of the potential.
Figures~\ref{fig:twoModeSpec}(c) and (d) show the same values as (a) and
(b) but for the Hamiltonian
\begin{equation}
  H = 
    \begin{pmatrix}
     \rmi \gamma(1+a_I) + a_R & -1\\
    -1 & -\rmi \gamma(1-a_I) -a_R 
    \end{pmatrix},
  \label{eq:TwoAsymHamiltonian}
\end{equation}
with $a_R = -0.15$ and $a_I = -0.2$.

Introducing an increased on-site energy in the loss and a decreased energy in
the gain well one gets rid of the exceptional point.
This can be understood intuitively since the probability densities in the two
wells now differ for the ground and excited state, i.e., either the gain or the
loss well is favored.
However, the two previously $\PT$-broken states and the intersection point are
not lost.
Instead, the ground state $\mu_-$ turns into the upper branch, intersects with
the axis $\imag \mu = 0$, and forms a stationary state.
Since the ground state has a larger probability of presence in the
energetically lower gain well, its out-coupling of particles is weakened and
the imaginary part of the chemical potential is shifted upwards.
Consequently, the intersection point $\gamma_0$ for $U = 0$ is shifted to a
lower parameter as compared to the case $a_R = 0$.

In this new configuration, a repulsive interaction tends to equalize both
densities.
This weakens the influence of the real asymmetry and shifts the imaginary part
of the chemical potential down to smaller values, thus, the intersection point
moves again to larger values of $\gamma$.
Now, as expected, the instability due to norm oscillations vanishes and no
Bogoliubov-de Gennes eigenvalue has a positive imaginary part.
In fact, all eigenvalues apart from the trivial solution $\omega = 0$
describing a phase shift have negative imaginary parts at the intersection
point, at which the stationary state with $\imag\mu = 0$ resides.
The stationary state is therefore not only stable, but acts as an attractor.
If the in- and out-coupling parameter is not at the intersection point
$\gamma_0$, the wave function's norm grows or decays to match the appropriate
interaction strength for which $\gamma_0 = \gamma$ holds.

It is clear that the intersection point for $U = 0$ is the lowest possible
parameter $\gamma$ for which such an attractor can be found.
If this lowest parameter is set, the wave function is attracted to the norm
$0$, i.e., the condensate will completely deplete.
Note that due to the strong particle out-coupling the same happens if even
smaller values of $\gamma$ are used.
To determine this threshold, the linear wave equation is solved, and one finds
that the eigenvalue of the ground state,
\begin{equation}
  \mu_- = \rmi \gamma a_I - \sqrt{1+(a_R+\rmi \gamma)^2},
  \label{eq:eigenvalueGroundStateFullAsym}
\end{equation}
becomes real for 
\begin{equation}
  \gamma = \frac{1-a_R^2(a_I^{-2}-1)}{1-a_I^2}.
  \label{eq:crossingPointLinear}
\end{equation}
This shows that the limit $\gamma = 0$ is reached for $a_R =
a_I/\sqrt{1-a_I^2}$.
Even stronger real asymmetries lead to a dominance of the gain contribution and
therefore to a completely unstable system.

One remark has to be made on the choice of the parameters $a_R$ and $U$.
The negative sign of the parameter $a_R$ leads to the intersection point of the
eigenvalue of the ground state with the axis $\imag\mu = 0$.
Obviously, a positive sign of the parameter would lower the particle
out-coupling of the excited state.
In this case an intersection of the eigenvalue and the axis $\imag\mu = 0$ can
be achieved for attractive interaction.
Therefore, it is possible to achieve real stationary states for both $U>0$ and
$U<0$.
However, a real ground state is only possible using repulsive interactions and
$a_R<0$.

\section{Dynamics and convergence}
\label{sec:twoModeDyn}
Each state of the non-Hermitian two-mode system is defined by three real
parameters,
\begin{equation}
  \psi(R,\phi,\theta) = R
  \begin{pmatrix}
    \cos{(\theta/2)} \rme^{-\rmi \phi/2}\\
    \sin{(\theta/2)} \rme^{+\rmi \phi/2}
  \end{pmatrix}.
  \label{eq:BlochSphere}
\end{equation}
Using the norm $R$ of the state as the radius, and the two angles $\theta \in
[0,\pi]$ and $\phi \in [0,2\pi)$ as spherical coordinates, every state $\psi$
is represented by a point in the three-dimensional real space.

Since the dynamics is not norm-conserving, it is not possible to restrict the
discussion to the surface $R = 1$, as it was possible for Graefe et
al.~\cite{Graefe08a} studying the dynamics
of the $\PT$-symmetric Bose-Hubbard dimer.
Instead, an analysis of the complete three-dimensional dynamics as known from
the $\PT$-symmetric double-well~\cite{Haag14b} is required.

\begin{figure}
  \centering
  \includegraphics{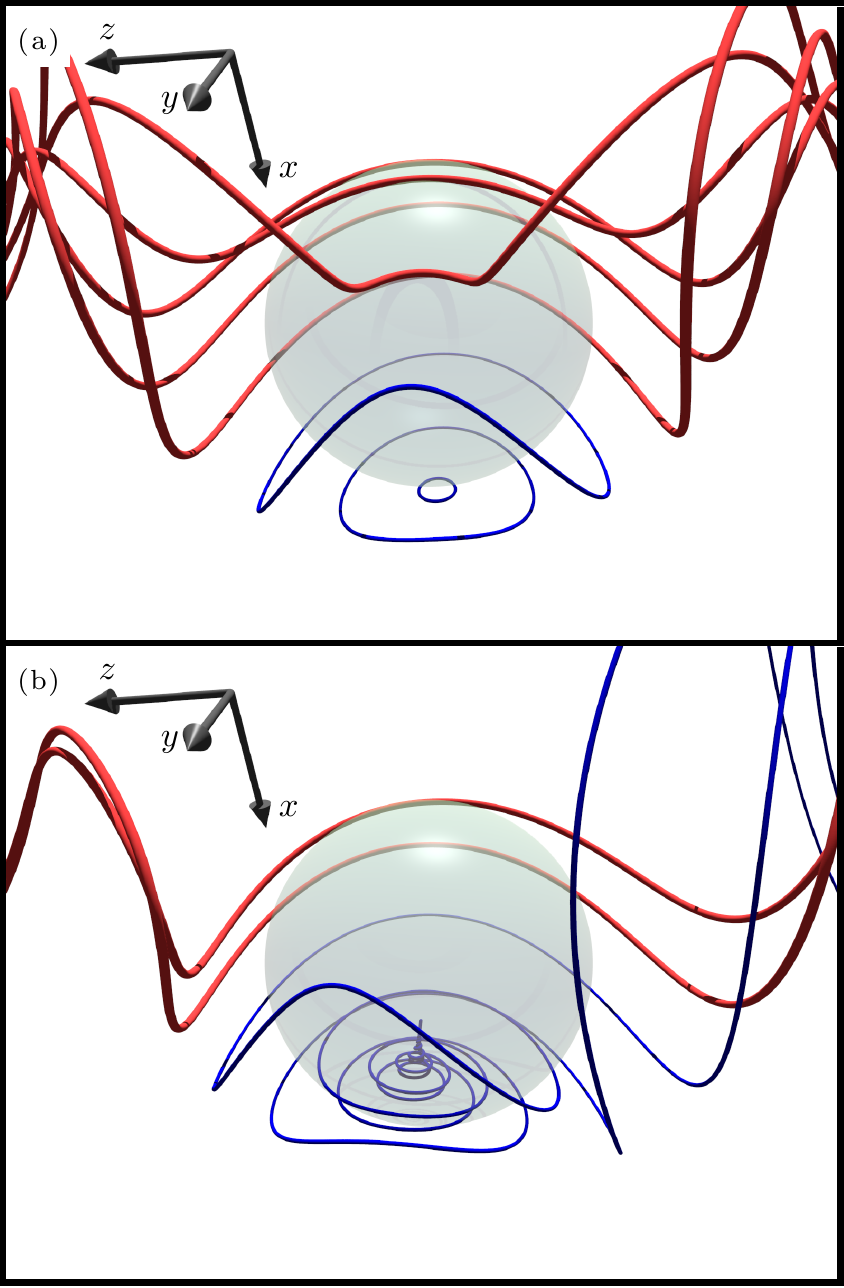}
  \caption{
    Bloch sphere for the parameters $U = 1$ and $\gamma = 0.7$.
    Since the radius of each state equals its norm, the transparent
    sphere at radius one represents all normalized states.
    Large $z$ values, which reside on the left side of the figure, correspond
    to states mainly residing in well one, i.e.\ the gain well, while the right
    side contains all states favoring the loss well.
    The $\PT$-symmetric sphere (a) shows stable oscillations (thin blue closed
    lines), while diverging trajectories (thick red open lines) start in the
    loss well for $t \to -\infty$ and end up in the gain well for $t \to
    \infty$.
    All trajectories are symmetric with respect to the $x$-$y$-plane.
    In the $\PT$-broken case (b) this symmetry is lost.
    While the upper two diverging trajectories (thick red lines) still run from
    the right to the left side, the converging trajectories (thin blue
    converging lines) are no longer closed.
  }
  \label{fig:twoModeBloch}    
\end{figure}

Figure~\ref{fig:twoModeBloch} shows this representation for both the
$\PT$-symmetric (a) and the asymmetric case (b) with $a_R = -0.15$, $a_I =
-0.2$ for the parameters $U = 1$ and $\gamma = 0.7$.
The dynamics of the $\PT$-symmetric system are defined by two types of
trajectories.
If the wave function lies near a stable stationary point (near the south pole
in Fig.~\ref{fig:twoModeBloch}(a)), it will start to oscillate around this
point forming a closed trajectory.
If it lies far away from such a stable fixed point, the wave function will
follow the $\PT$-broken state, while increasing its norm to
infinity~\cite{Haag14b}.

Studying Fig.~\ref{fig:twoModeBloch}(b) one immediately notices major
differences.
A dynamical attractor exists near the south pole approximately where the
original stable ground state of the $\PT$-symmetric system resides.
It lies at a norm smaller than one and at $z > 0$, favoring the gain well.
This is the only stable stationary point in Hilbert space and every trajectory
must either converge to this point or diverge to a state with a large norm
mainly localized in the gain well.
Due to this convergent norm oscillations, which are one of the characteristic
features of $\PT$-symmetric systems, can only be observed for a few
oscillations.
The amplitude of such a norm oscillation decays until the appropriate norm,
i.e., the appropriate particle number of the stationary state, is reached.

\begin{figure}
  \centering
  \includegraphics{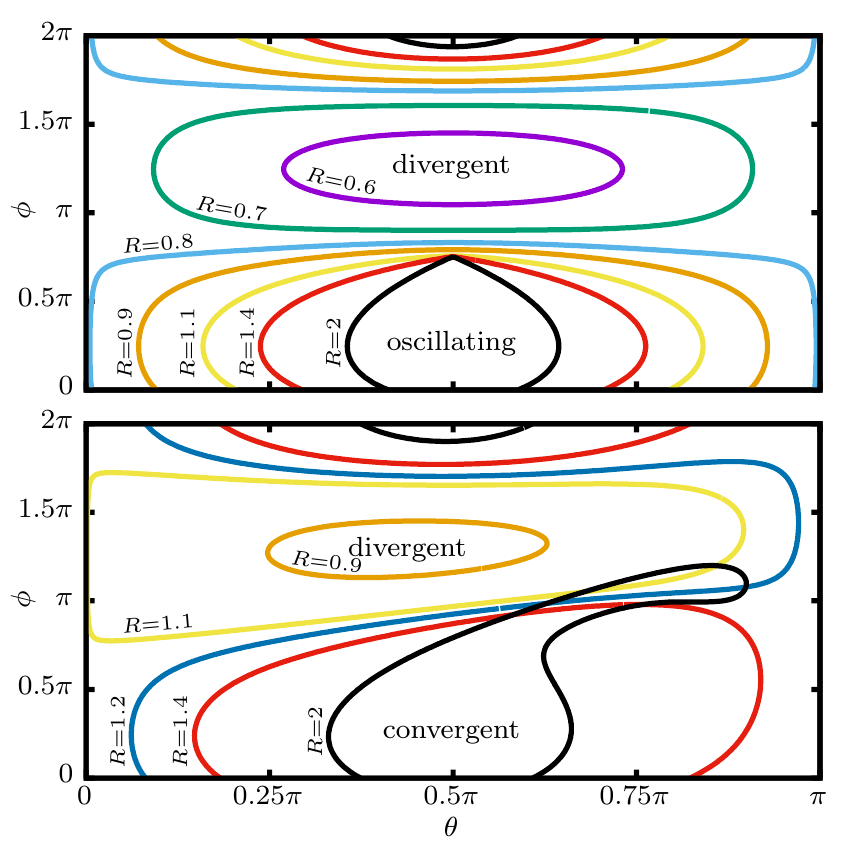}
  \caption{
    The intersection between the separatrix and spheres with given radii in the
    three-dimensional Bloch space for $U = 1$ and $\gamma = 0.7$.
    The upper panel shows the $\PT$-symmetric case with $a_R = 0$ and $a_I =
    0$ where the separatrix divides the space between stable oscillations and
    diverging wave functions.
    The lower panel presents the asymmetric case with $a_R = -0.15$ and $a_I =
    -0.2$; due to the attractor, no oscillating trajectories can be found and
    all wave functions are either divergent or convergent.
    In both cases, the separatrix reaches the lowest radii for $\theta \approx
    0.5\pi$ and $\phi \approx 1.25 \pi$.
    For higher radii, more wave functions become divergent, until only a small
    region around $\theta \approx 0.5 \pi$ and $\phi \approx 0.25\pi$ remains
    stable.
    The two cases differ in two aspects:
    The asymmetric case supports convergent wave functions with higher norms
    and is not symmetric to the axis $\theta = 0.5 \pi$ like the
    $\PT$-symmetric situation.
  }
  \label{fig:twoModeSeparatrix}    
\end{figure}

It is apparent that the convergent area of the asymmetric system is similar to
the area of stable oscillations in the $\PT$-symmetric case.
The only qualitative difference between these two cases lies in the fact that
the region of the $\PT$-symmetric case is closed in positive and negative
$z$-direction, while its counterpart with $a_R = -0.15$ and $a_I = -0.2$
includes strongly asymmetric states with large norms from the loss well.
The exact separatrix between the divergent and the attractive region can be
calculated in a straightforward manner, however, a three dimensional
presentation as done in Fig.~\ref{fig:twoModeBloch} does not allow a
quantitative analysis.
Instead, it is beneficial to restrict the calculation to discrete interaction
strengths, i.e., discrete radii in the Bloch representation, which can be
characterized as either convergent or divergent.
The results are shown in Fig.~\ref{fig:twoModeSeparatrix}, in which the
$\PT$-symmetric and the asymmetric case are compared.
In both cases, the first configurations that start to diverge are evenly
distributed between both wells and possess currents from the loss to the
gain well due to the phase difference $\phi \approx 1.25 \pi$.
At this point the first important difference becomes apparent.
While the $\PT$-symmetric case becomes unstable at $R \approx 0.6$, the
asymmetric case does not show any diverging trajectories up to $R \approx
0.85$.
For larger radii the divergent region grows quickly, symmetric around $\theta =
0.5 \pi$ for the $\PT$-symmetric case and strongly asymmetric for the case $a_R
= -0.15$ and $a_I = -0.2$.
The asymmetry is the second important difference between both cases.
This is essential for large radii, as the case $R = 2$ demonstrates in
Fig.~\ref{fig:twoModeSeparatrix}.
Due to the broken symmetry wave functions with a major occupation in the
loss well are able to converge to the attractor.

\section{Extended potentials}
\label{sec:extended}
In the previous sections we discussed a two-mode system which describes a
Bose-Einstein condensate with asymmetric gain and loss of particles.
For a more quantitative analysis the one-dimensional Gross-Pitaevskii equation
is solved numerically exact without the restriction to a finite set of basis
vectors.
The Gross-Pitaevskii equation reads
\begin{equation}
\left(  - \partial _x ^2 + V(x) + g |\psi(x,t)|^2 \right) \psi(x,t) = \mrm{i}
\partial _t \psi(x,t)
\label{eq:ExtendedGPE}
\end{equation}
with the strength of the nonlinear contact interaction $g$ and the asymmetric
complex double-well potential
\begin{equation}
  V(x) = 
  \begin{cases}
    V_1(x),\quad x \leq 0 \\
    V_2(x),\quad x > 0
  \end{cases}.
  \label{eq:eq:AsymmetricPotentialCases}
\end{equation}
The left half of the potential, $x \leq 0$, is chosen as
\begin{equation}
V_1(x) = 
\frac{1}{4} x^2  + 4 e^{- \frac{1}{2} x^2} - i \gamma  x e^{-0.12 x^2}.
\label{eq:AsymmetricPotentialxgtr0}
\end{equation}
It describes a single well composed of a harmonic trapping potential and a
Gaussian barrier at $x = 0$. 

\begin{figure}
  \centering
  \includegraphics{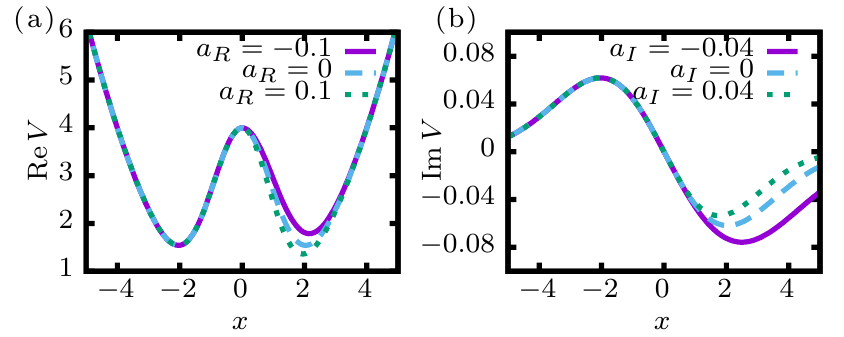}
  \caption{Real part (a) and  imaginary part (b) of the spatially extended
    potential~\eqref{eq:AsymmetricPotential} and
    \eqref{eq:AsymmetricPotentialxgtr0} for various asymmetry parameters and
    $\gamma = 0.05$.
  }
  \label{fig:AsymmetricPotential}    
\end{figure}

The imaginary part is positive and thus describes particle gain.
Its strength is controlled  by the parameter $\gamma$.
The right well, $x > 0$, is modified,
\begin{equation}
  V_2(x) = \frac{1}{4}  x^2  + 4  e^{- (\frac{1}{2} + a_R)  x^2} 
          - i \gamma  x e^{(-0.12 + a_I) x^2}.
\label{eq:AsymmetricPotential}
\end{equation}
In this well, particle loss is applied which is weakened for $a_I > 0$ and
strengthened for $-0.12 < a_I < 0$.
The parameter $a_R$ shrinks or expands the barrier, i.e., lowers or raises the
potential in the loss well.
Note that, as in the two-mode system, negative values of $a_R$ and $a_I$
correspond to a stronger particle loss and a shallower loss well.
For the case $a_R = a_I = 0$, both halves become equal and result in an
extensively studied $\PT$-symmetric potential~\cite{Dast13a, Dast13b, Haag14b}.
The potential is shown in Fig.~\ref{fig:AsymmetricPotential} for various
parameters $a_R$ and $a_I$.

\begin{figure}
  \centering
  \includegraphics{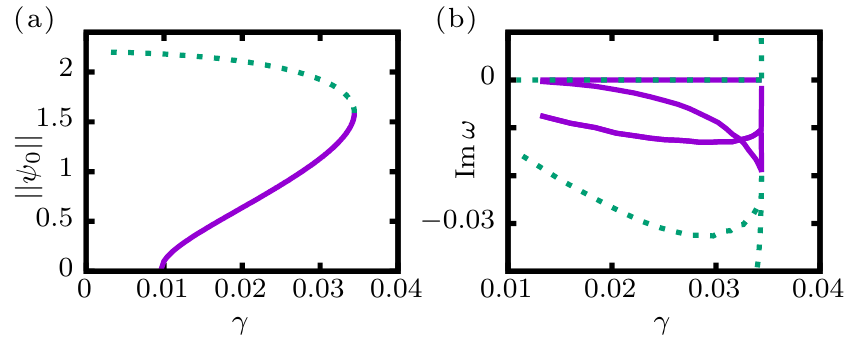}
  \caption{
    Norm of the stationary states (a) and their four smallest Bogoliubov-de
    Gennes eigenvalues (b) over the value of the in- and out-coupling parameter
    $\gamma$ for $g = 0.1$, $a_R = -0.01$ and $a_I = -0.08$.
    As long as larger values of $\gamma$ support a higher norm of the
    stationary state (solid line), it is stable.
    Therefore, the upper branch (dotted line) is unstable for all parameters
    $\gamma$.
  }
  \label{fig:RealEigenvaluesInteractionStrength}    
\end{figure}

First we study whether and where stable stationary states can be found for
non-vanishing asymmetries.
Numerical calculations show that the conclusions from the two-mode system hold
true also for the spatially extended potential.
The ground state shows the desired behavior for repulsive interactions and a
stronger particle loss if, at the same time, the on-site energy in the loss
well is increased.
We therefore set the parameters to $a_R = -0.01$, $a_I = -0.08$, and $g = 0.1$.
The parameters used correspond to a relative imbalance of the well's depths of
about $1\%$ and an imbalance of the particle in- and out-coupling of about
$25\%$.
Note that, compared to the experiment~\cite{Gati07a}, the interaction strength
is very weak, i.e., Feshbach resonances~\cite{Inouye98a} would have to be
employed.
The stationary state for each parameter $\gamma$ is calculated and its norm,
i.e., the necessary nonlinear strength to stabilize this in- and out-coupling,
is shown in Fig.~\ref{fig:RealEigenvaluesInteractionStrength}(a).
For most parameters $\gamma$ two stationary states can be found.
The lower state in Fig.~\ref{fig:RealEigenvaluesInteractionStrength}(a) has a
higher stable norm for larger values of $\gamma$, thus, stabilizing
itself against norm oscillations as discussed previously for the two-mode
model.
The second branch results from the fact that stronger interaction strengths
shift the stationary state back to smaller parameters $\gamma$.
This was shown in Fig.~\ref{fig:twoModeSpec}(c).
However, in this range the state is not protected against norm oscillations,
and therefore is unstable.

The Bogoliubov-de Gennes eigenvalues shown in
Fig.~\ref{fig:RealEigenvaluesInteractionStrength}(b) confirm these
considerations.
There are three different types of perturbations, one of which is the trivial
solution with $\imag \omega = 0$.
The second type becomes zero for a vanishing norm.
Additional studies of the data confirm that the eigenvalue is approximately
linear in the squared norm of the stationary state and negative.
This is exactly the behavior we expect from a stable perturbation of the wave
function's norm and is only found for the lower branch.
The eigenvalues of the third type have a finite value even for a vanishing
norm.
At this point, the particle-particle interaction vanishes, i.e., the
perturbation exists even for linear systems.
It corresponds to a higher excited state which decays exponentially since the
particle out-coupling is stronger than the in-coupling into the system.
Even though only one of them is shown in
Fig.~\ref{fig:RealEigenvaluesInteractionStrength}, all such eigenvalues are
negative, i.e., they correspond to stable solutions.

To put the quantitative results into perspective, we shortly remind of the
corresponding results from the $\PT$-symmetric system~\cite{Dast13a}.
For small parameters $\gamma$, this system supports two stationary states for a
wide range of interaction strengths.
The $\PT$ symmetry breaking occurs at $\gamma \approx 0.042$ after which the
$\PT$-symmetric stationary states vanish.
Figure~\ref{fig:RealEigenvaluesInteractionStrength} shows that the range of
stable stationary ground states is $0.01 \lesssim \gamma \lesssim 0.035$, i.e.,
it includes most of the original range in the $\PT$-symmetric system.

As a final test the attractive behavior of the stationary state is examined in
dynamical calculations.
To reduce the number of parameters describing the time-dependent wave function, 
the Bloch-sphere representation is used again.
Since the Hilbert space is not two-dimensional anymore, we choose a projection
onto the space spanned by the normalized ground state $e_1 =  \psi _g$ and the
orthogonal vector $e_2 = \alpha \left( \psi _e - \langle \psi _g, \psi
_e\rangle \psi _g \right)$ selected by the Gram-Schmidt method, where
$\alpha$ is the normalization constant.
Calculations from \cite{Haag14b} for the $\PT$-symmetric case show that this is
a good approximation for the system considered.
An arbitrary state in this basis can be written as in
Eq.~\eqref{eq:BlochSphere} and is therefore defined by the three real
parameters $R$, $\phi$ and $\theta$.
Note that the orthonormal two-dimensional basis is different from the choice
made in the two-mode approximation and, thus, the parameters $\phi$ and
$\theta$ must differ.
\begin{figure}
  \centering
  \includegraphics{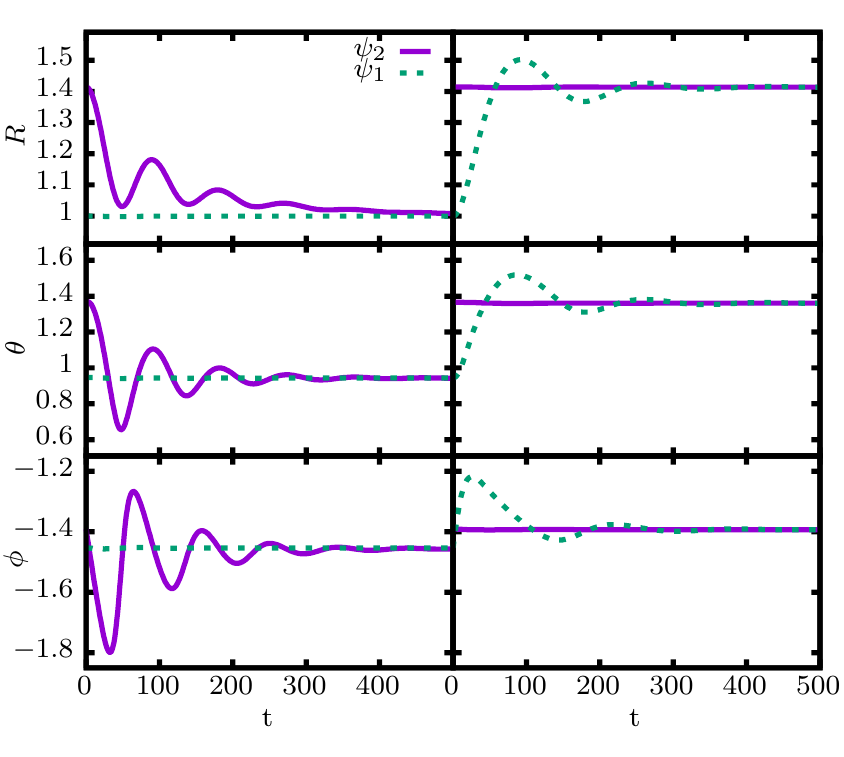}
  \caption{
    Time evolution of the Bloch coordinates of the two stationary states
    $\psi_i$ for $\gamma_1 \approx 0.0277$ (green dotted line) and $\gamma_2
    \approx 0.0366$ (magenta solid line).
    In the left panel a system with $g = 0.1$, $a_R = -0.01$, $a_I =
    -0.08$, and $\gamma \approx 0.0277$ is used.
    Hence, $\psi_1$ is stationary while $\psi_2$ converges to $\psi_1$ within
    approximately four oscillations.
    In the right panel the in- and out-coupling parameter is changed to 
    $\gamma \approx 0.0366$, and the two states exchange their roles.
  }
  \label{fig:TimeEvolutionExtended}    
\end{figure}

We study two distinct cases which differ only in the strength of the in- and
out-coupling parameter $\gamma_1 \approx 0.0277 $ and $\gamma_2 \approx
0.0366$.
Both cases support a stationary state $\psi_i$ with norms $R_1 = 1$ and $R_2 =
\sqrt{2}$ and the two Bloch angles $\phi_i$ and $\theta_i$.
Figure~\ref{fig:TimeEvolutionExtended} shows the time evolution of the three
Bloch coordinates for both states in both systems.
We see that the coordinates $R_i$, $\phi_i$, $\theta_i$ of both states converge
to the appropriate dynamical attractor, i.e., they converge to the coordinates
of $\psi_1$ in the left panel and of $\psi_2$ in the right panel.
The time of the convergence to the stationary states differs in the two cases.
This results from the difference of the smallest Bogoliubov-de Gennes
eigenvalue which is inversely proportional to the timescale of the convergence.
Here $|\imag \omega |$ from $\psi _2$ is larger than $\psi _1$, therefore, the
time of convergence for $\psi _1$ is smaller.
This is confirmed by Fig.~\ref{fig:TimeEvolutionExtended}.
Thus, we have shown that the behavior of the two-mode system is also found in a
realistic spatially extended system, i.e., for a large range of values of the
coupling strength $\gamma$, the stationary states act as attractors.

\section{Conclusion}
In this work we studied Bose-Einstein condensates in an asymmetric
non-Hermitian double well.
In an experiment such asymmetries are always unavoidable.
However, we were able to show that it is possible to manipulate the system in
such a way that an attractor of the dynamics exists which possesses all
properties of an $\PT$-symmetric state required to identify it in an
experiment.
First, we presented stationary solutions to the Gross-Pitaevskii equation of a
two-mode system with asymmetric gain and loss.
One requirement for a stable realization can be formulated a priori:
If the particle gain is stronger than the particle loss, perturbations with
high excitation energies will be exponentially enhanced rendering the state
unstable.
Therefore, the particle loss must always be stronger than the gain.
In this configuration, the ground state of the system cannot only be made
stable but becomes a dynamical attractor.
To achieve this, a real asymmetry of the trapping potential was introduced,
reducing the particle density of the ground state in the loss well, i.e., the
stronger particle loss is partially counterbalanced by the asymmetric trap.

Next, the dynamical properties were carefully studied using a specific
asymmetric potential and a fixed repulsive contact interaction.
Weak asymmetries leave the $\PT$-symmetric oscillations mainly intact.
However, all such wave functions end up at the dynamical attractor, effectively
limiting the timescale during which such oscillations can be observed.
It was shown that the convergent region is indeed even larger than in the
$\PT$-symmetric case including a set of wave functions with large norms
residing in the loss well.

Finally, the results were compared to those of a realistic spatially extended
potential.
Not only do all observations from the two-mode approximation remain valid, but
in addition we were able to show that an attractor exists in a wide range of
particle in- and out-coupling strengths.
Therefore, the potential is capable of acting as a setup for a $\PT$-symmetric
realization with unbalanced gain and loss.

\end{document}